\documentclass[modern,trackchanges]{aastex62}
\usepackage{amssymb, amsmath}
\usepackage{color}
\usepackage{morefloats}
\usepackage{hyperref}

\shorttitle{A Runaway Yellow Bright Giant in the Galactic Halo}

\shortauthors{Massey et al.}

\NewPageAfterKeywords

\begin{document}

\title{A Runaway Giant in the Galactic Halo\footnote{This paper includes data gathered with the 6.5 meter Magellan Telescopes located at Las Campanas Observatory, Chile.}}

\author{Philip Massey}
\affiliation{Lowell Observatory, 1400 W Mars Hill Road, Flagstaff, AZ 86001, USA}
\affiliation{Dept.\ of Physics and Astronomy, Northern Arizona University, Flagstaff, AZ, 86011-6010, USA}
\email{phil.massey@lowell.edu }

\author{Stephen E. Levine}
\affiliation{Lowell Observatory, 1400 W Mars Hill Road, Flagstaff, AZ 86001, USA}
\affiliation{Department of Earth, Atmospheric and Planetary Sciences, MIT, Cambridge, MA 02139, USA}

\author{Kathryn F. Neugent}
\affiliation{Department of Astronomy, University of Washington, Seattle, WA, 98195, USA}
\affiliation{Lowell Observatory, 1400 W Mars Hill Road, Flagstaff, AZ 86001, USA}

\author{Emily Levesque}
\affiliation{Department of Astronomy, University of Washington, Seattle, WA, 98195, USA}

\author{Nidia Morrell}
\affiliation{Las Campanas Observatory, Carnegie Observatories, Casilla 601, La Serena, Chile}

\author{Brian Skiff}
\affiliation{Lowell Observatory, 1400 W Mars Hill Road, Flagstaff, AZ 86001, USA}

\begin{abstract}
New evidence provided by the {\it Gaia} satellite places the location of the runaway star J01020100-7122208 in the halo of the Milky Way (MW) rather than in the Small Magellanic Cloud as previously thought.  We conduct a reanalysis of the star's physical and kinematic properties, which indicates that the star may be an even more extraordinary find than previously reported. The star is a 180~Myr old 3-4$M_\odot$ G5-8 bright giant, with an effective temperature of 4800$\pm$100~K, a metallicity of ${\rm Fe/H}=-0.5$, and a luminosity of $\log L/L_\odot=2.70\pm0.20$.  A comparison with evolutionary tracks identifies the star as being in a giant or early asymptotic giant branch stage. The proper motion, combined with the previously known radial velocity, yields a total Galactocentric space velocity of 296~km~s$^{-1}$.   The star is currently located 6.4~kpc below the plane of the Milky Way, but our analysis of its orbit shows it passed through the disk $\sim$25~Myr ago.  The star's metallicity and age argue against it being native to the halo, and we suggest that the star was likely ejected from the disk. We discuss several ejection mechanisms, and conclude that the most likely scenario is ejection by the Milky Way's central black hole based upon our analysis of the star's orbit.  The identification of the large radial velocity of J01020100-7122208 came about as a happenstance of it being seen in projection with the SMC, and we suggest that many similar objects may be revealed in {\it Gaia} data.

\end{abstract}

\section{Introduction}

The second data release (DR2) of the European Space Agency (ESA) satellite {\it Gaia} is revolutionizing stellar and Galactic studies thanks to the abundance of high precision parallax and and proper motion information \citep{DR2}.  In addition to the plethora of new discoveries this enables, the DR2 data have revealed inadvertent errors in interpretation of previous discoveries. For instance, these data have shown that many of the late-type low-mass stars previously thought to be hypervelocity stars (HVSs) are actually bound to the Milky Way, with their velocities previously overestimated \citep{GaiaHVS}.  In other cases, the {\it Gaia} data reveal that objects we thought were interesting may not be quite what we thought, but may prove to be just as odd and exciting.  We discuss one such case here. 

We recently reported follow-up observations of a star thought to be a runaway yellow supergiant in the Small Magellanic Cloud (SMC) \citep{YSGRunaway}.  The star, J01020100-7122208 (UCAC2 1342515), had been previously shown by \citet{NeugentSMC} to have a heliocentric radial velocity of $\sim$300 km s$^{-1}$, considerably greater than the velocity expected for its projected position in the SMC, about 170~km~s$^{-1}$.  \citet{YSGRunaway} described new spectroscopic observations which confirmed the large velocity, as well as establishing a spectral subtype of G5-8 with an effective temperature of 4700$\pm250$ K.  Their conclusion was this was a yellow supergiant runaway, possibly ejected at high speed as a result of a binary companion exploding as a supernova, in accord with the original scenario suggested for runaways by \citet{Zwicky} and \citet{Blaauw}.

However, DR2 shows this star has a parallax of $0.0728\pm0.0190$ mas, far too large for an SMC member, and placing  J01020100-7122208 in the inner halo of the Milky Way.  Simply inverting this parallax leads to a value of 13.8~kpc, but as most DR2 users have discovered, this is not an accurate process for converting a parallax to a distance when the uncertainties are a significant fraction of the value.  In addition, there is a small systematic error in the {\it Gaia} parallaxes, in the sense that they are too small, typically by $\sim$0.03~mas \citep{Lindegren}.  However, this offset may vary globally by as much as 0.1~mas, depending upon position in the sky.  An examination of the quasars behind the SMC by \citet{NGC362} suggests that this $-0.03$~mas average offset  is appropriate for the region where our star is located.   Applying this offset, and using the {\it Gaia} Python software\footnote{https://repos.cosmos.esa.int/socci/projects/GAIA/repos/astrometry-inference-tutorials/browse}, we find a median distance modulus of 14.75~mag (8.9~kpc) adopting an exponentially decreasing space density, the least biased estimator found by \citet{Gaiaparallaxes}.    The 90\% confidence interval (excluding any uncertainty in the systematic error) extends from a distance modulus of 14.26~mag (7.1~kpc) to 15.34~mag (11.7~kpc). We adopt $14.8\pm0.5$~mag for the distance modulus\footnote{In addition to establishing the changes in the parallax zero point with position in the sky,  the {\it Gaia} quasar parallaxes also suggest there {\it may} be variations in the zero-point of a similar size depending upon magnitude and color.   As emphasized by \citet{Lindegren}, it is also not clear at this time what the implications are for the uncertainties for brighter sources.  Our giant has a {\it Gaia} G magnitude of 13.4, while most of the quasar sample have $G>17$.  Including an additional 0.03~mas uncertainty in the parallax would increase the uncertainty on the distance modulus from $\pm0.5$~mag (the formal uncertainty) to $\pm0.9$~mag, causing an uncertainty on the star's luminosity of $\pm0.36$~dex. We have not carried this additional error through since it is not clear that is a realistic value, or even exists,  but will remind the reader of this at appropriate junctures.}.  The SMC is located at a distance modulus of 18.9~mag (59~kpc), well outside this range.    Further,  the measured proper motion ($8.647\pm0.036$ mas yr$^{-1}$ in right ascension, and $-0.906\pm0.027$ mas yr$^{-1}$ in declination) are an order of magnitude larger than what we would expect if the star were located in the SMC and its transverse motion were similar to its radial velocity. 

\begin{figure}
\epsscale{1}
\plottwo{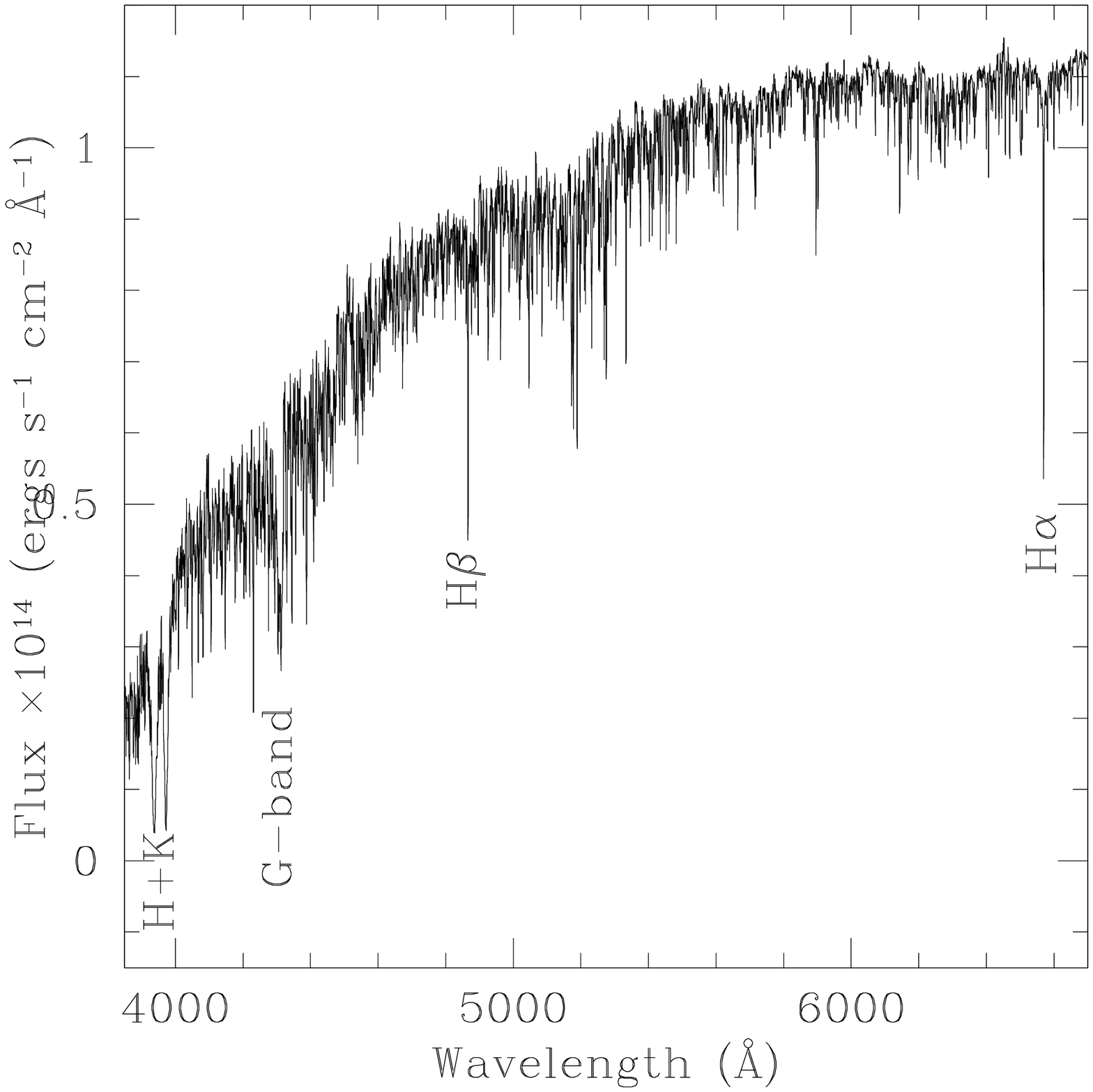}{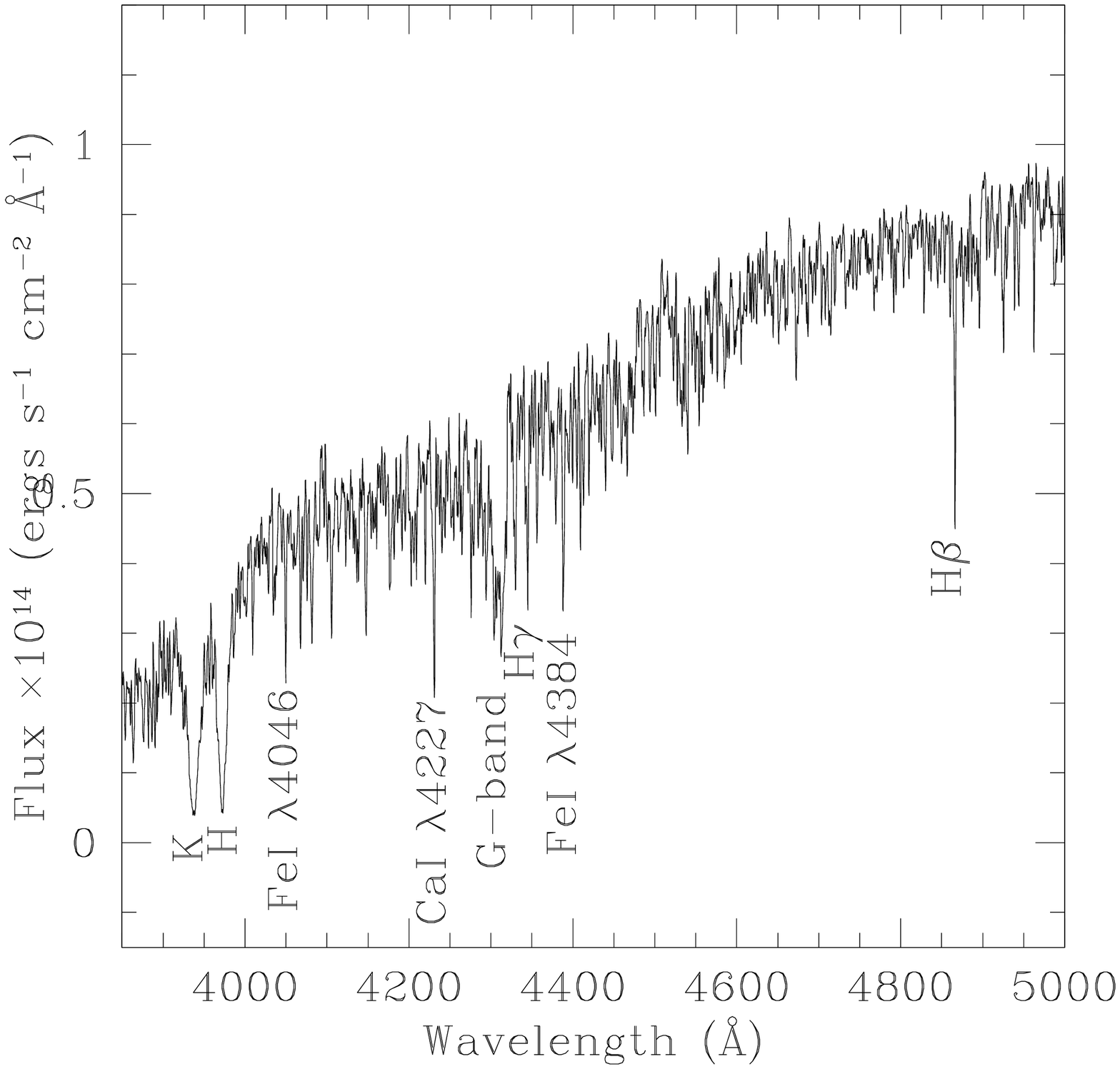}
\caption{\label{fig:fluxedspectrum}  The fluxed spectrum of J01020100-7122208.  We show the flux-calibrated spectrum of our star, based upon the data described in \citet{YSGRunaway}.  The prominent G-band, strengths of the Ca\,{\sc ii} H and K lines, and the strengths of the Balmer lines indicate a G5-8~II spectral type.  The plethora of metal lines is apparent, showing that the star is not a metal-poor poor halo giant.}
\end{figure}

In our original paper, we briefly considered the possibility that our star was a halo giant, but rejected this argument for what seemed at the time to be a valid reason.  Although some halo giants are known to have large radial velocities, as a population they are very metal poor, with [Fe/H] $\le -1.5$ (see, e.g., \citealt{Helmi}). The spectrum of J01020100-7122208 shows a prominent G-band, and an abundance of metal lines (Figure~\ref{fig:fluxedspectrum}).  Furthermore, the widths of the H and K Ca~II lines demonstrated that the star is of relatively high luminosity.   

In this followup paper, we derive new physical properties in Section~\ref{Sec-prop}, present the kinematics in Section~\ref{Sec-Kinematics}, discuss the implications and possible origins of this star in Section~\ref{Sec-Discussion}, and summarize our results in Section~\ref{Sec-summary}.

\section{Physical Properties and Location in the Hertzprung-Russell  Diagram (HRD)}
\label{Sec-prop}

What, then, is this star, and is it still of astrophysical interest?  First, let us reexamine the physical properties of the star (luminosity, metallicity, effective temperature, and age) given the new information about its distance.   

\citet{YSGRunaway} compared the spectrum of the star against spectral standards, finding that it was G5-8.  The star has an apparent visual magnitude of $V=13.75$ \citep{DennisSMC}.  Adopting a distance modulus of $14.8\pm0.5$ as indicated by the {\it Gaia} parallax (see above), leads to an absolute visual magnitude $M_V=-1.1$ in the absence of any reddening, considerably brighter than that of a typical G5-8 giant ($M_V=+0.9$, \citealt{Allen}), more like a luminosity class II bright giant.   This is consistent with the evidence presented in \citet{YSGRunaway} that the H and K lines indicate high luminosity.  Accounting for extinction would result in the star being more luminous than this value.
 
\begin{figure}
\epsscale{1}
\plotone{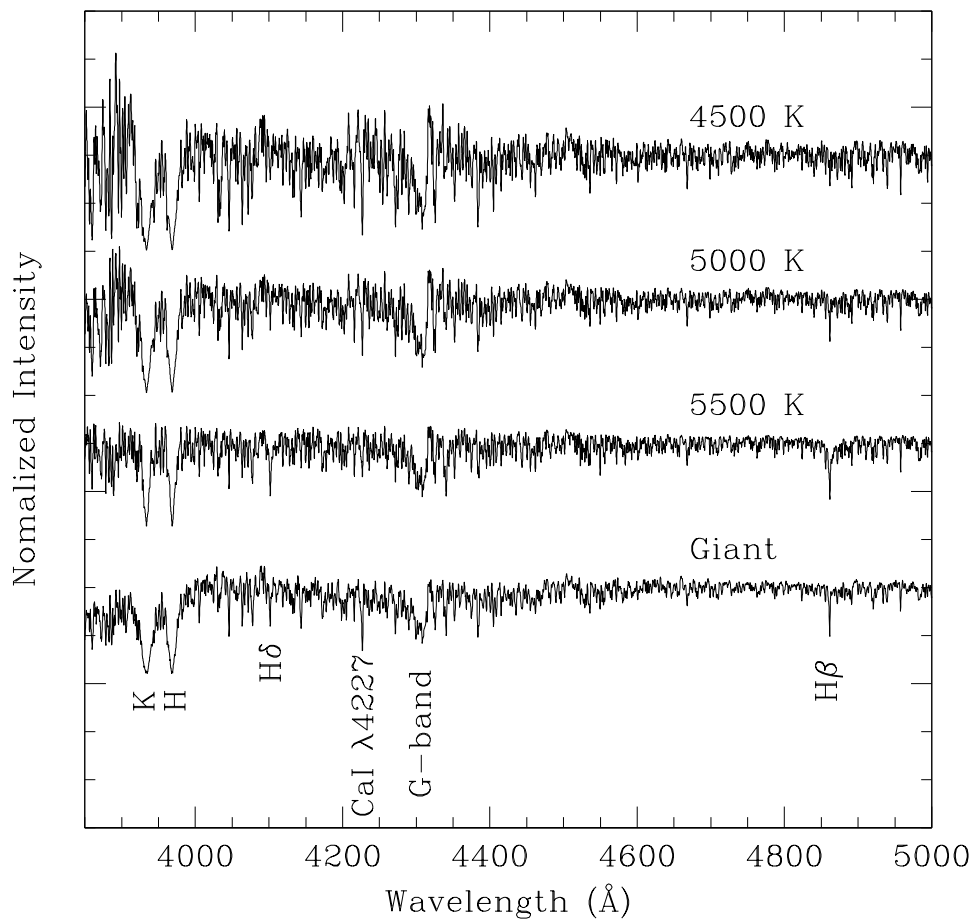}
\caption{\label{fig:spectrum}  The normalized spectrum of J01020100-7122208.  We have compared the spectrum of our star with that of three PHOENIX models with effective temperatures of 4500~K, 5000~K, and 5500~K. The strengths of the H and K Ca\,{\sc ii} lines, the Ca\,{\sc i} $\lambda 4227$ lines, the G-band, and the Balmer lines indicate a temperature near 5000~K, consistent with what we expect from the spectral type, and with the {\it Gaia} temperature determination, and that expected simply from the star's spectral type.}
\end{figure}

We expect an effective temperature of 4750-5050~K for a G5-8~II star \citep{Allen, Gerard, 1999A&AS..140..261A}.  The {\it Gaia}-derived temperature is 4867~K.  This value is based upon the Gaia broad-band photometry using a data-driven algorithm, which attempts to separate the effects of reddening from temperature; the formal uncertainty  in this regime is 324~K \citep{GaiaTemps}.  ({\it Gaia} extinction is derived separately and is currently unavailable for our star, due to the parallax uncertainty being greater than 20\%.)  In  Figure~\ref{fig:spectrum} we compare the normalized spectrum of our star to three PHOENIX models \citep{Phoenix} of various temperatures.  The strengths of the Balmer lines and weakness of the G-band rules out a temperature as low as 4500~K, while the strength of the Ca\,{\sc i} $\lambda 4227$ line suggests a temperature below 5000~K.  This is in good agreement with the  {\it Gaia} value, as well as what we expect based upon the spectral classification. 
The bolometric correction corresponding to this range temperature is $\sim$$-0.3$~mag \citep{Kurucz04}, and hence we can set a lower limit on the bolometric luminosity of the star: $M_{\rm bol} < -1.4$, or $\log L/L_\odot >2.4$. This is a lower limit as we have made no correction for interstellar extinction; the actual value must be larger.   

\begin{figure}
\epsscale{1}
\plotone{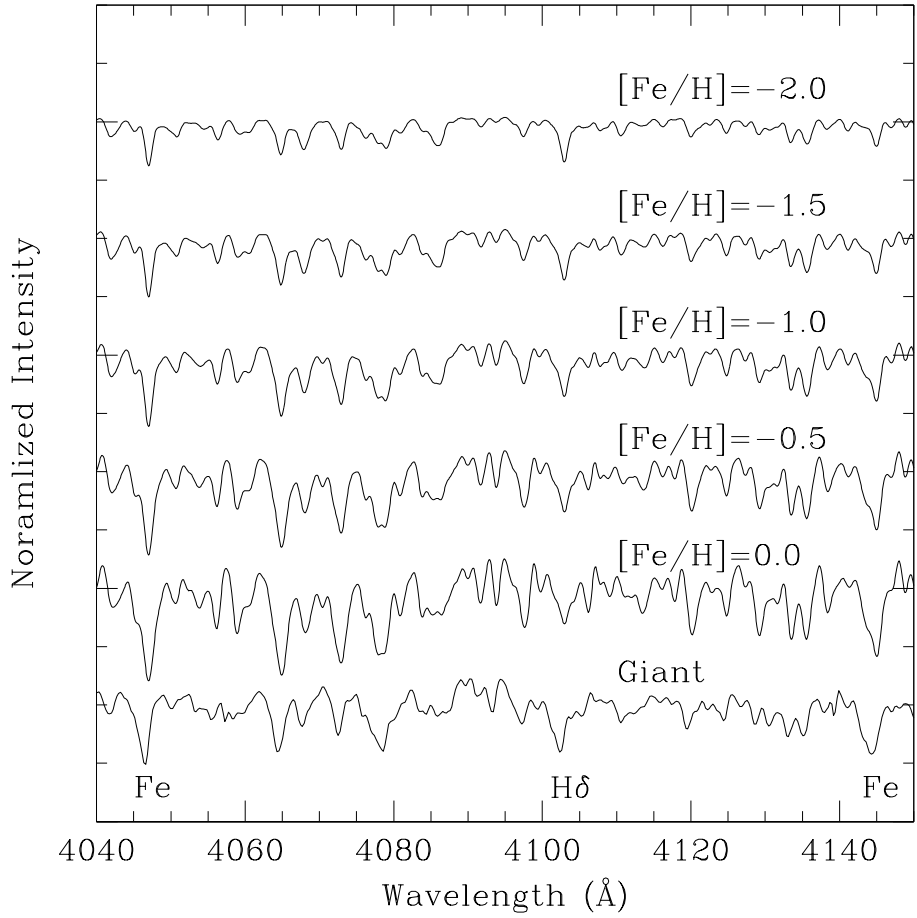}
\caption{\label{fig:metal} Metallicity of our giant. The spectrum of our giant star is compared to PHOENIX models with metallicity of [Fe/H]=0.0 to -2.0.  The strengths of the Fe\,{\sc i} $\lambda$4045 and Fe\,{\sc i} $\lambda$ 4143 lines with respect to H$\delta$ suggest a metallicity of around -0.5. The spectrum of our star comes from \citet{YSGRunaway} with a resolution of $R$ of 4100, and the models are all 4800~K with $\log g$[cgs] = 2.0, and come from \citet{Phoenix}.}
\end{figure}

Using this estimate of the temperature, we can now use the PHOENIX models to determine the metallicity.   As long as the temperature is fairly well known, the ratios of Fe\,{\sc i} $\lambda$4143 to H$\delta$ and Fe\,{\sc i} $\lambda$4045 to H$\delta$ give a good indication of the approximate metallicity.  In Figure~\ref{fig:metal} we compare the normalized 4800~K Phoenix models of various [Fe/H] metallicities  to the spectrum of our giant.  (We have used the $\log g=2.0$ models as these correspond to the approximate value we derive below; the method is insensitive to the exact choice of surface gravities.)  The strengths of the Fe lines allow us to exclude the possibility that our star has a metallicity as low as $-1.0$, or as high as solar, and we adopt
a value of [Fe/H] of $-0.5$.

To refine the effective temperature, we next turn to the broad-band photometry; this will also allow us to calculate the extinction, needed for a better estimate of the star's luminosity.  We use the ATLAS models \citep{Kurucz04} as these have broad-band colors readily available on-line\footnote{http://wwwuser.oats.inaf.it/castelli/colors/bcp.html}. This is an iterative process, as we must adopt a surface gravity, which depends upon the adopted mass, which depends upon the luminosity.  Fortunately, the dependence of the intrinsic model temperature is also not very sensitive to the exact value of the surface gravity: in our temperature---$\log g$ regime, the change in $(B-V)_o$  is about 0.05~mag over 1~dex in $\log g$, comparable to the photometric error.  We estimate a mass of roughly 3$M_\odot$ from the Geneva evolutionary tracks of \citet{Sylvia} and the MESA tracks of \citet{MIST}, which leads to a surface gravity $\log g\sim 2.2$.  We adopt 2.0 provisionally, as we know the luminosity (and hence the radius) will be larger, and thus the surface gravity a bit lower.
In terms of the observed photometry, we adopt the optical {\it UBV} values from \citet{DennisSMC}, and the 2MASS values for the near-IR \citep{2MASS}.  There is only one $U-B$ value in the literature, but there are multiple sources for the other colors, all of which are in agreement within the uncertainties.

In Table~\ref{tab:red} we list the observed photometry and derived extinction $A_V$, using the relationships $E(U-B)=0.72E(B-V)$, $E(V-K)=2.95E(B-V)$, and $E(J-K)=0.54E(B-V)$, where the first is the standard value, and the later two come from \citet{Sch}.  We have adopted a value for the total-to-selective extinction $R_V=3.1$. 

\begin{figure}
\epsscale{1}
\plotone{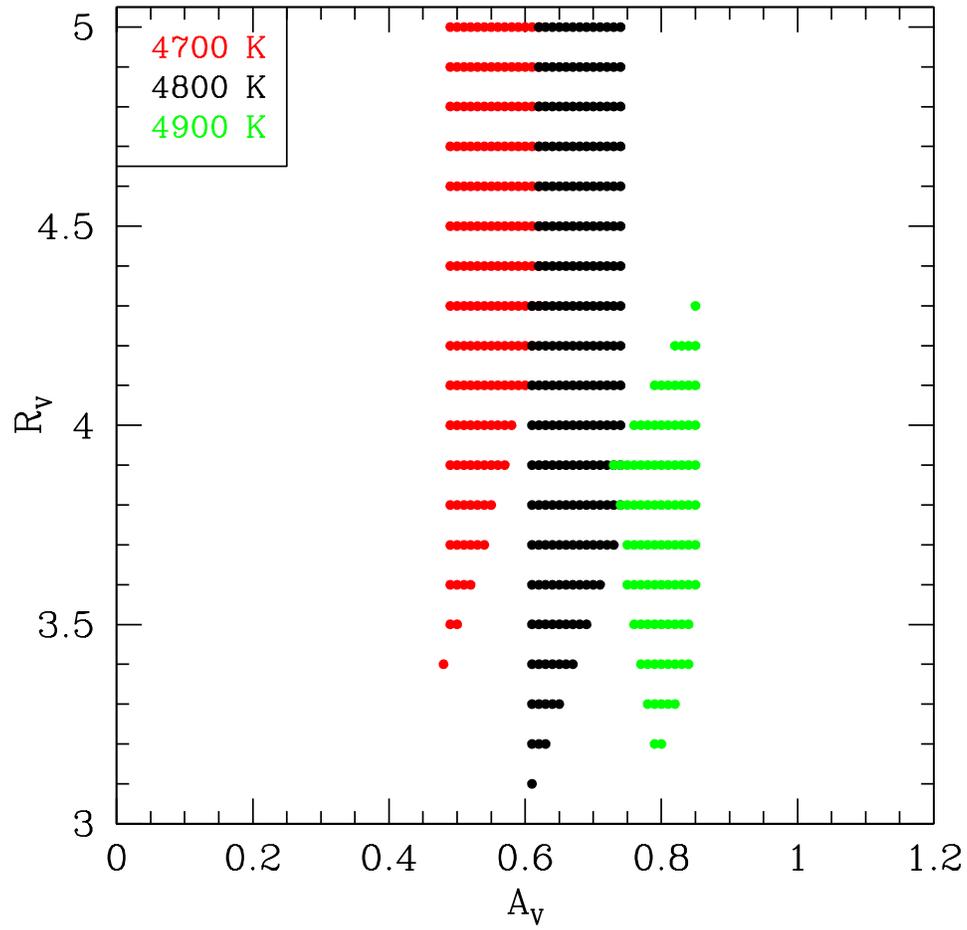}
\caption{\label{fig:avr} Allowed ranges of $R_V$ and $A_V$.  In order for the \citet{CCM} reddening law to produce consistent values of $A_V$ for the four colors $U-B, B-V, V-K,$ and $J-K$, the $R_V$ and $A_V$ values must be in the ranges indicated as a function of temperature. No valid values were found for temperatures at 4600~K and below, or for 5000~K and above.}
\end{figure}

The agreement in $A_V$ between the various color indices is poorer than what we would expect.  To remove this inconsistency, we explore the effect of changing the ratio of total to selective extinction $R_V$ from its canonical 3.1 value, adopting the \citet{CCM} reddening law.  We look for consistency within the errors for each trial temperature, $R_V$ value, and $A_V$ value.  We show the result in Figure~\ref{fig:avr}.  We included lower and higher temperatures in our exploration, but found no regions where the $A_V$ values agreed for 4600~K and cooler, for 5000~K and above.   It seems likely that $R_V$ is at least slightly higher than 3.1, and we adopt a value of 3.5.  The likely $A_V$ is $0.65\pm0.15$~mag.  (This uncertainty is still small compared to that of the distance modulus, $\pm$0.5~mag.)
Note that this extinction is considerably higher than what we expect.  The total foreground reddening in the direction of the SMC is $A_V=0.12$ based upon the  \citet{Sch} 100-$\mu$m dust emission maps; see Table~17 in \citet{LGGSII}.  

We thus adopt an effective temperature of 4800$\pm100$~K, consistent with the spectral type, {\it Gaia}  temperature, and the optical and NIR photometry.  The absolute visual magnitude is $M_V=-1.7\pm0.5$.  The bolometric luminosity is
$M_{\rm bol}=-2.0\pm0.5$, or $\log L/L_\odot=2.70\pm0.20$.  (If we allow for the full range of possible zero point errors in the
{\it Gaia} parallaxes, the luminosity is $\log L/L_\odot =2.70\pm0.56$) The stellar radius is 32$\pm$8$R_\odot$.

\begin{figure}
\epsscale{0.8}
\plotone{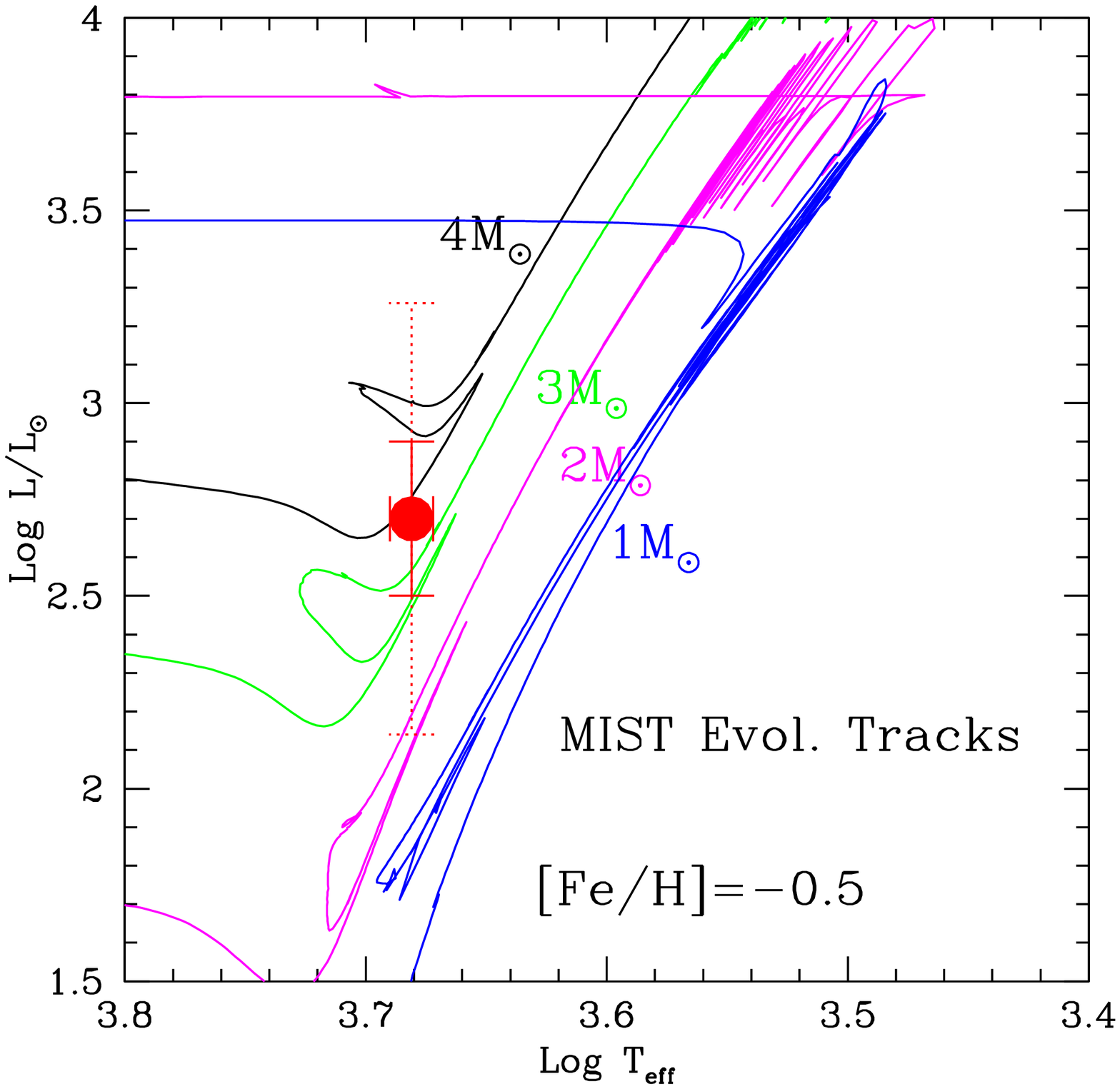}
\plotone{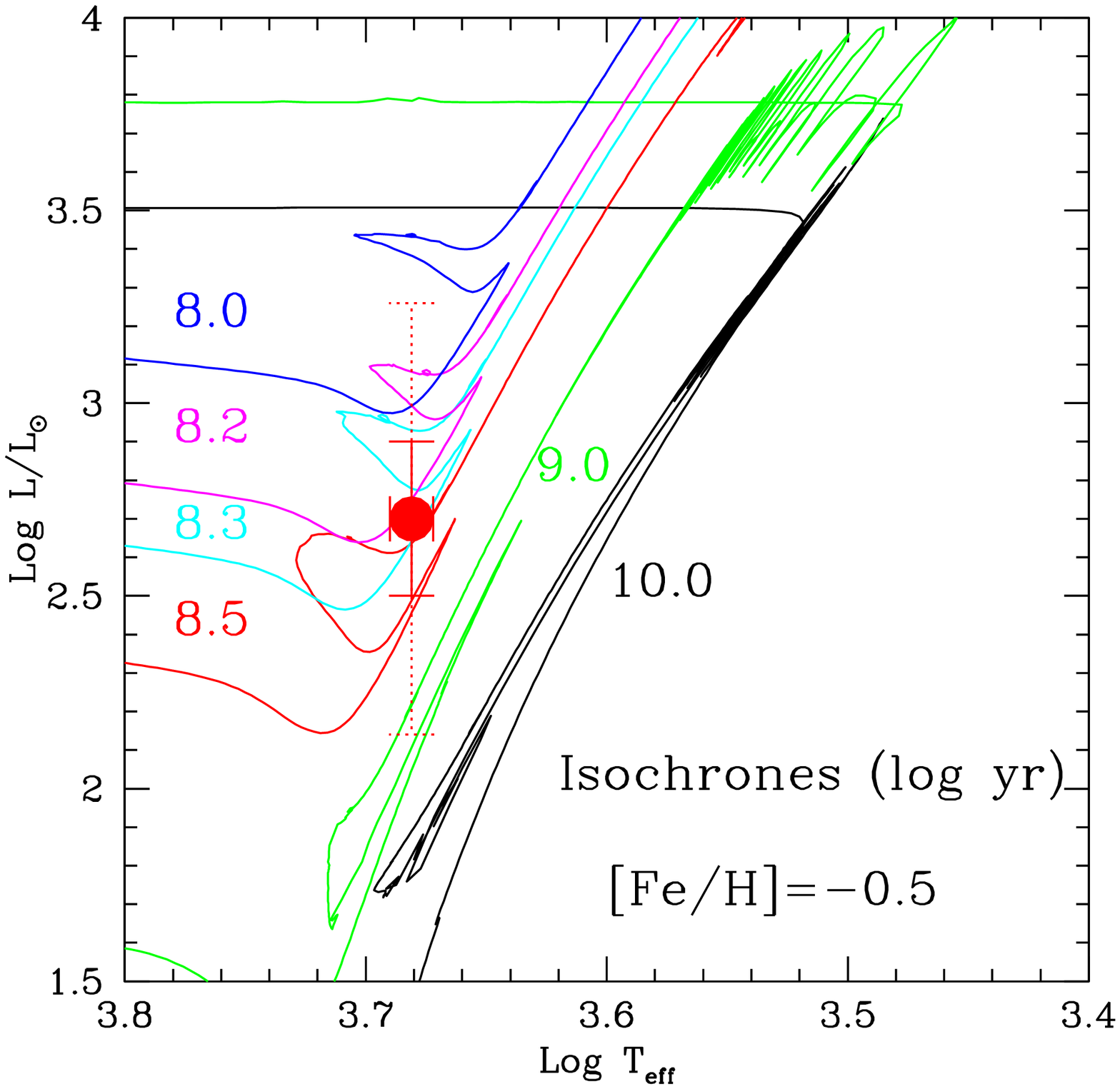}
\caption{\label{fig:HRD} {\it Top:} Position in the H-R Diagram.  The MIST \citep{MIST} evolutionary tracks are shown for a metallicity [Fe/H]=-0.5).  The position of our star is indicated, along with the error bars.  The dotted extension of the luminosity error bar corresponds to the full potential zero-point error of the {\it Gaia} parallax. {\it Bottom:} The corresponding isochrones are shown.}
\end{figure}

The physical parameters we have arrived at are listed in Table~\ref{tab:physical}, and the star's location in the H-R diagram is shown in Figure~\ref{fig:HRD} ({\it top}).   The star has a likely mass between 3 and 4$M_\odot$. The star is on the red giant branch or early asymptotic branch (AGB) stage.  The  $\sim$0.5~mag of extra extinction we find around the star (0.65~mag vs.\ the 0.12~mag we expect) is thus easily explained, as such stars are expected to lose about 0.5$M_\odot$ during the red giant phase due to radiation pressure on dust and radial pulsations (see, e.g., \citealt{LamersLevesque}, p. 16-8, and references therein), much of it in the form of dust.   A larger-than-average $R_V$ value implies a distribution of grain sizes that is larger than normal,  as is commonly seen for circumstellar dust around ``smokey" stars, such as red supergiants (see, e.g., discussion in \citealt{Smoke}; some extreme examples are given in \citealt{1987ApJ...321..921S} and \citealt{2015A&A...584L..10S}).  The MIST isochrones are shown in Figure~\ref{fig:HRD} ({\it bottom}).  The star has an age of about 180~Myr, with a likely range of 125-300~Myr.  

This is not an age typical of the halo. Note that even if the most pessimistic version of the errors (shown by a dotted line in Figure~\ref{fig:HRD}) the age of our star is constrained to be less than 1~Gyr (log age $<$ 9), not the $>10$~Gyr (log age = 10) that characterize the halo population (see, e.g., \citealt{Helmi,Jofre}).  At the same time, the metallicity of our star is much higher than found among halo stars. Note the role played here by the star's spectrum, which not only determines the star's metallicity, but fixes the key observational parameter, namely the star's temperature.  For instance, the possibility that this star is actually a low-mass AGB star, say, can be excluded on the basis of the spectrum alone. Figure~\ref{fig:HRD} shows that a 1$M_\odot$ AGB star would have a  $\log T_{\rm eff} \sim 3.62$ (4100~K), a temperature that would correspond to a late K-type giant.  (That said, it would also not be allowed by the color, as such a star would be redder than what we observe by several tenths in $B-V$.)

\section{Kinematics}
\label{Sec-Kinematics}

We initially grew interested in this star because of its high radial velocity.  Now that the star's proper motion and distance are known, the star's kinematics is, if anything, even more interesting. 

The heliocentric radial velocity measured by \citet{YSGRunaway} is 301$\pm2.4$~km~s$^{-1}$.  The proper motion measured by {\it Gaia} are $\mu_\alpha \cos \delta = 8.647\pm0.036$ mas yr$^{-1}$ and $\mu_\delta=-0.906\pm0.027$~mas yr$^{-1}$.  At a distance of 8.9~kpc, this transverse velocity would correspond to 367~km~s$^{-1}$ with respect to the sun.

It is more meaningful to discuss this in terms of Galactocentric coordinates and velocities.  We adopt a distance from the sun to the Galactic center of 8.5~kpc.  The Cartesian $xyz$ position of the star with respect to the Galactic center is
$x=-5.2$~kpc, $y=-5.3$~kpc, and $z=-6.4$~kpc, where the $x$ axis is defined from the sun ($x=-8.5$~kpc) to the Galactic center ($x=0$), and the $xyz$ is a right-handed coordinate system, with the $z$ component out of the plane of the Milky Way.  Thus the star is (currently) located well beyond the thick disk, in the halo.   The Cartesian velocity with respect to the local standard of rest (LSR) in the standard UVW plane (with U positive in an outwards direction) is 165~km~s$^{-1}$, $-400$~km~s$^{-1}$, and $-167$~km~s$^{-1}$, where we have adopted a solar motion of (U,V,W) of $-10$~km~s$^{-1}$, 5.2~km~s$^{-1}$,  7.2 km s$^{-1}$ \citep{1998MNRAS.298..387D}.  Relative to the Galactic center (i.e., with the circular motion of 220~km~s$^{-1}$ of the LSR removed; see \citealt{1986MNRAS.221.1023K}), this converts to Cartesian velocities of $X=-165$~km~s$^{-1}$, $Y=-180$~km~s$^{-1}$, and $Z=-167$~km~s$^{-1}$.  

\begin{figure}
\epsscale{1}
\plottwo{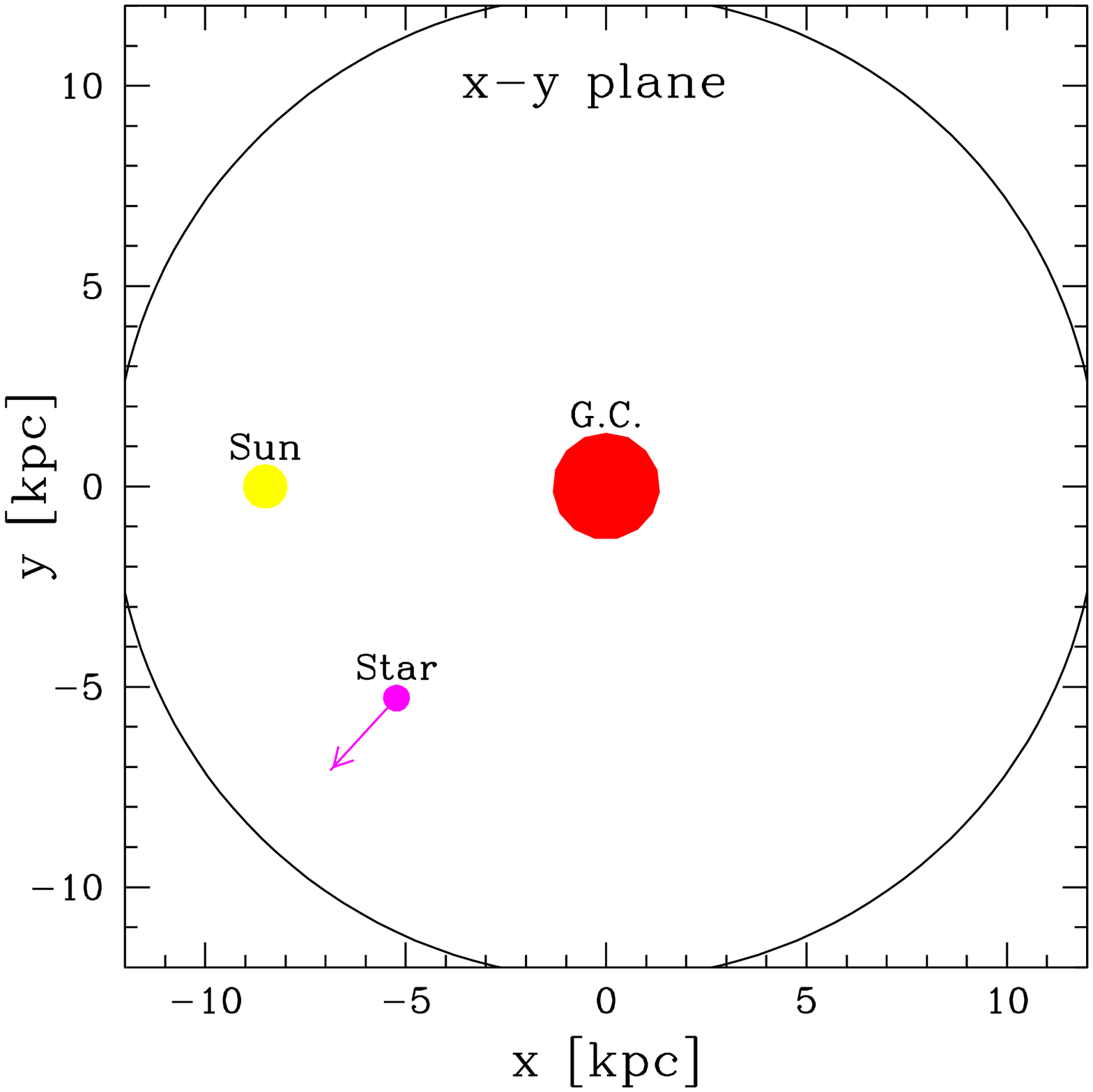}{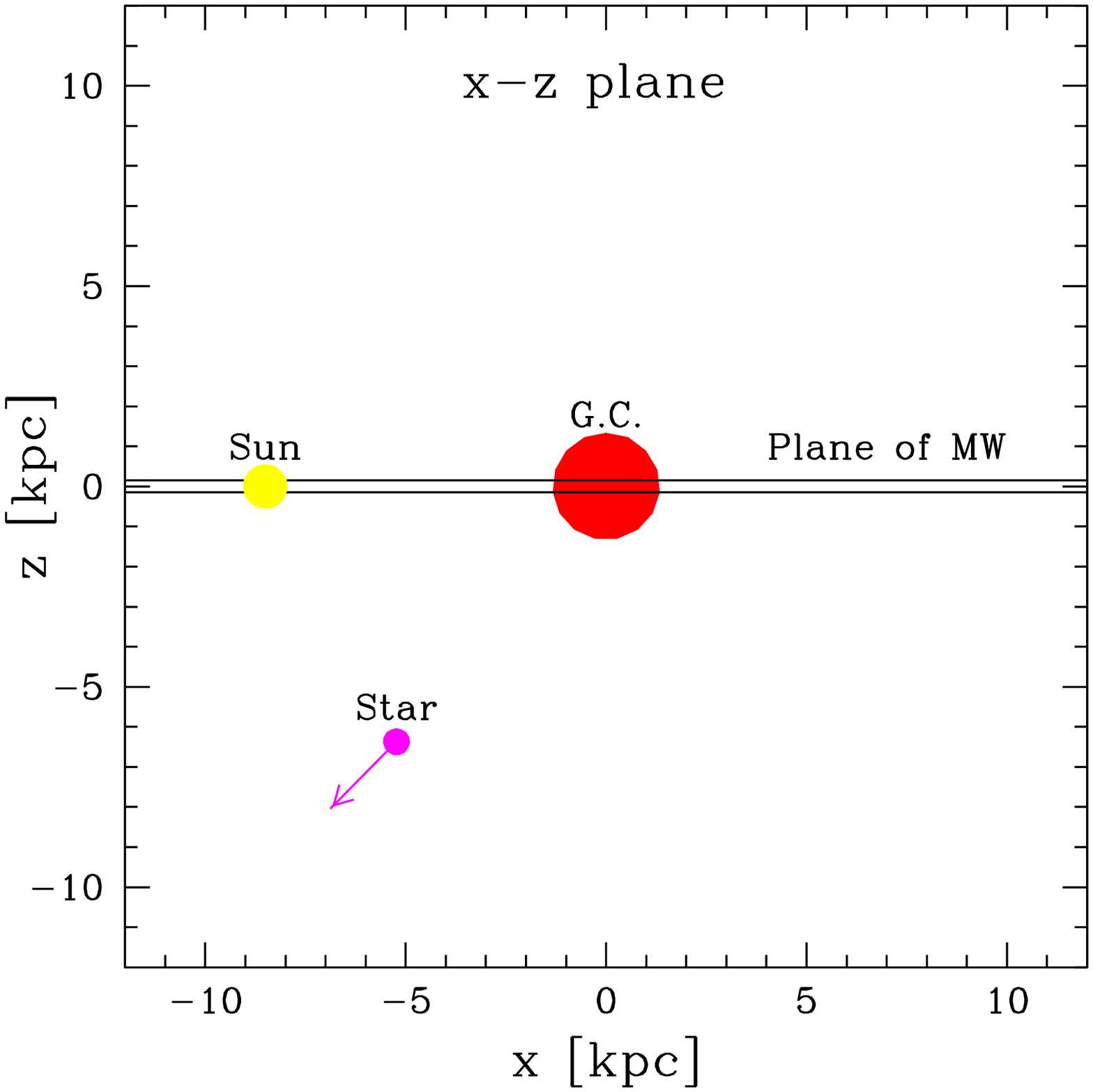}
\epsscale{0.5}
\plotone{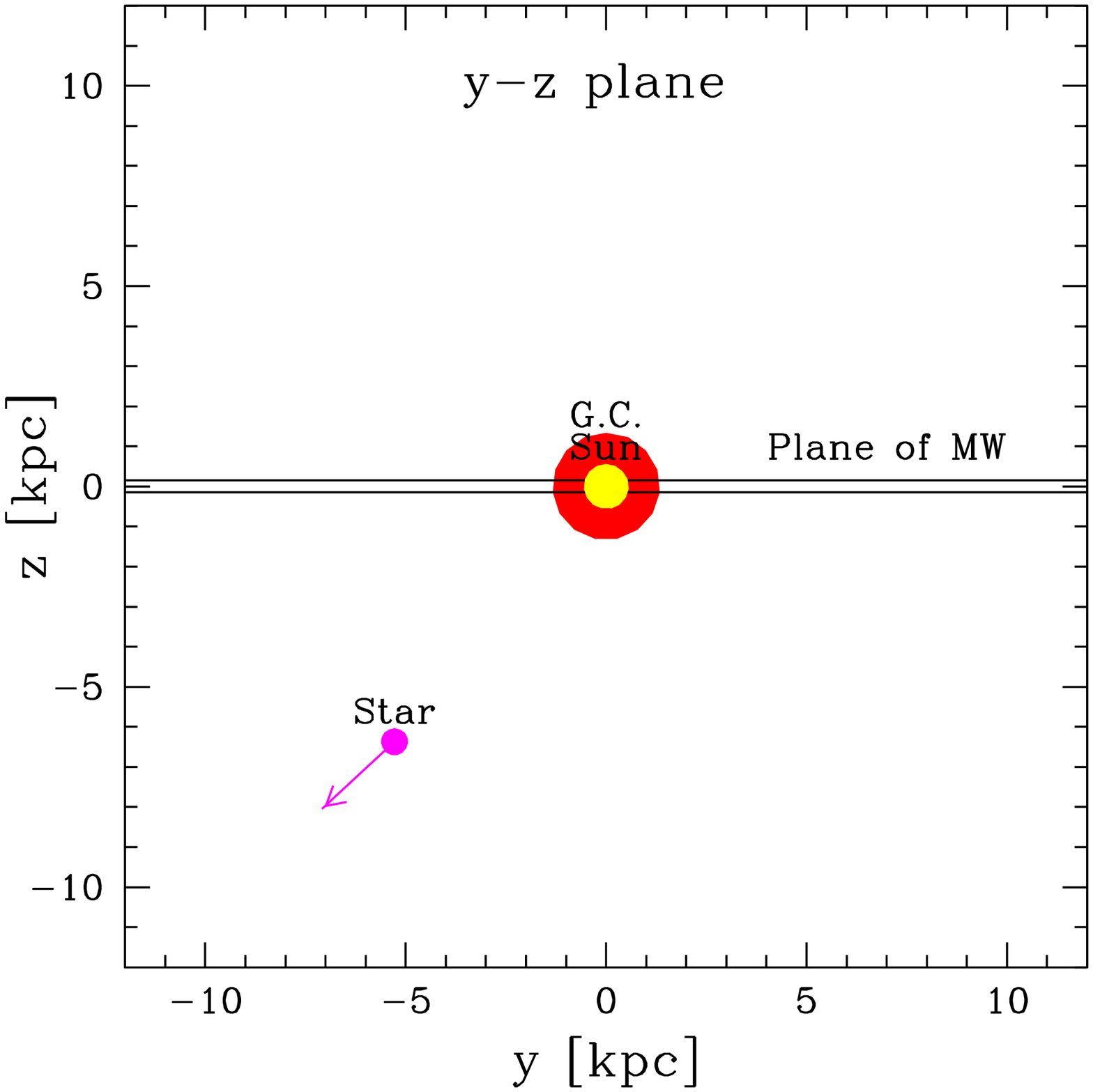}
\caption{\label{fig:loc}  Position and velocity of J01020100-7122208 in Galactocentric coordinates.  The location of the star is shown with respect to the Galactic center.  The vectors show the velocities in terms of the tangential motion in 10~Myr.}
\end{figure}

We summarize the kinematics in Table~\ref{tab:kinematics}, and in Figure~\ref{fig:loc} we show the location of the star with respect to the Galactic center (G. C.).  The vectors show the current direction and velocities of the star.

If we think of this in terms of spherical coordinates we find that the the star's orbit is primarily radial: the galactocentric radial distance between the star and the Galactic center $R$ is 9.8~kpc.  Its galactocentric radial velocity is about 293 km s$^{-1}$ while its Galacrocentric tangential velocity is about 34~km~s$^{-1}$.  (The total space velocity is thus about 296 km s$^{-1}$.)

Having the full phase space coordinates in hand, we undertook to
integrate the possible orbit of this star backwards in time. We
followed the orbit in four possible cases. The first two models
represented extremes of the possible Galaxy's mass distribution (a
Keplerian point mass potential, and a pseudo-isothermal sphere). The
third model is a quasi-realistic approximation based on the work of
\citet{1995A&A...300..117D}. This uses three Miyamoto-Nagai \citep{1975PASJ...27..533M}
components to represent the bulge, disk and halo of the
Galaxy. The last model was a variant of the third where we added in a
massive black hole (MBH) at the center of the Galaxy.

We integrated the star's orbit using a Bulirsh-Stoer integrator (see, e.g., \citealt{NR}, Section 17.3).
The orbits were integrated backwards for $5 \times 10^8 \, {\rm yr}$. The
phase space data were dumped every 100,000 years. Energy in the
simulations was conserved to better than a part in $10^9$.

In all four models, the
star passed through the galactic plane between 22 and 27~Myr
years ago. The orbital period (assuming the star could live so long)
is between 240 and 300~Myr (if bound); thus the star has made at most one pass through the Galactic plane.  The
perigalacticon distances range between 130 and 780 pc for all four
models. The apogalacticon distances are roughly 23~kpc, except for
the Keplerian model, where the star is unbound. At perigalacticon
passage, the maximum total velocity ranged between 451~km~s$^{-1}$
for the pseudo-isothermal model up to 2,576~km~s$^{-1}$ for the Keplerian. 
The Dauphole \& Colin models peaked at
726~km~s$^{-1}$. The orbit has an inclination with respect to
the galactic plane of about 78$^\circ$ and an eccentricity between 0.93 and 0.99
for the bound cases.

The primary reason for including the fourth model with the central MBH
 was to see whether that could be a scattering source for
this star. The two Dauphole \& Colin models (without and with the
MBH) are essentially identical, which is not too surprising given
that the closest approach distance is 130-780~pc for all four models (but see below). The orbit is shown in
Figure~\ref{fig:SEL}. The black square represents the star today, and the open
triangles were placed at 25~Myr intervals to show its past locations.  
To keep this in context, the star's current age is about 180~Myr, i.e., about 7 triangles;
earlier than that the orbit is shown as dots.

\section{Origins}
\label{Sec-Discussion}

J01020100-7122208 is not a typical halo giant.  Its temperature, combined with its visual brightness and  {\it Gaia} distance, shows that it is an intermediate-mass object; our analysis suggests a mass of 3.5$\pm0.5 M_\odot$, with an age of 180~Myr, not the old, metal-poor population that we usually associate with the halo.  The metallicity is [Fe/H]$\sim$$-0.5$, considerably higher than the $\leq-1.5$ that is typical of the halo.  Although not qualifying as a HVS, the total space motion (348~km~s$^{-1}$) of our star is quite high compared to the $\sim 100-150$~km~s$^{-1}$ velocity dispersion for halo stars with $R<15$~kpc \citep{2015ApJ...813...89K}\footnote{Although reliable proper motions for halo stars have not been available prior to {\it Gaia}, by undertaking wide-field radial velocity studies in carefully selected directions, studies such as \citet{2015ApJ...813...89K} have been able to determine the three dimension velocity dispersions of halo stars as a function of Galactocentric radius.}.

\subsection{Disk Origins}

The star's age suggests an origin in the disk---it is not a multiple-Gyr old object, and hence was not born as a halo object.  The orbit analysis shows the star was in the disk 22-27~Myr ago.   If J01020100-7122208 was ejected into the halo at that time, what are the likely mechanisms?

\subsubsection{Encounter with the Massive Black Hole in the Center of the Galaxy}

Dozens of high-velocity, late-type B stars have been found in the halo \citep{2015ARA&A..53...15B}.  These hypervelocity stars (HVSs) are traveling so fast that they are unbound.  They have masses of 2.5-4$M_\odot$\citep{2014ApJ...787...89B}, very similar to that of our star.  The prevailing belief is that they were ejected into the halo as a result of encounters with the MBH in the center of the Milky Way, as no other mechanism seems capable of explaining their numbers and extraordinary velocities.  In this scenario, a binary is disrupted by a close passage with the MBH due to tidal forces, with one component then bound to the MBH and the other ejected at velocities up to several thousand~km~s$^{-1}$\citep{1988Natur.331..687H}.

Could such an encounter have happened with our star?  The orbit is almost purely radial in Galactocentric coordinates, as shown above.  This strongly suggests that whatever scattered the star out of the plane was located at or near the center of the Galaxy.  The obvious culprit would be the central MBH.  Our initial orbit calculation, described above, does not seem to support this, as the closest approach is still hundreds of parsecs.  The HVSs must have passed within a small fraction of a parsec to have been ejected; see Equation 1 in \citet{2015ARA&A..53...15B}.  We were curious to see what would happen if we relaxed any of the input parameters to our orbit calculation.  It became apparent that the parallax uncertainty had the most effect upon our results; the other uncertainties (such as the transverse components and radial velocities) have little influence. We ran orbits for assumed distances of 8.3~kpc and 9.5~kpc to compare to our baseline assumption of 8.9~kpc. The results are shown in Figure~\ref{fig:rot}. 

The main thing to note is that the sense of rotation changes between the 8.9~kpc (Figure~\ref{fig:rot}b) and 9.5~kpc (Figure~\ref{fig:rot}c) models. Assuming a smooth potential etc, then a continuous function must cross the center for some value of the parallax between 8.9~kpc and 9.5~kpc.  {\it Thus there is a range of parallaxes for which the star would pass sufficiently close to the MBH to have been scattered into the halo!} This window is probably very small, but it must be there.  Given how radial the orbit is, this scenario seems probable.  Recall that the 90\% confidence interval for the distance is 7.1~kpc to 11.7~kpc; a value intermediate between 8.9 and 9.5 is well within the uncertainty.

Of course, future refinement in the parallax from {\it Gaia} could exclude this result.  However, with the present data ejection into the halo by the MBH seems highly likely.

\subsubsection{Dynamical Ejection from a Cluster}
A commonly invoked alternative explanation to the supernova scenario for the origins of OB runaway stars is the dynamical ejection from a cluster.  As reviewed by \citet{2012ApJ...751..133P}, a star is accelerated by passing close to one component of a hard binary.  (A ``hard" binary is one whose binding energy is greater than the average kinetic energy of single stars.) Some simulations have explored binary/binary interactions. For this to have any statistical likelihood, this needs to take place in a dense, compact cluster. By passing within the semi-major axis of the binary, the single star obtains
a kick that is comparable to the binding energy of the binary.   Indeed, some studies (e.g., \citealt{2009MNRAS.396..570G}) have shown that such encounters can readily eject 4$M_\odot$ stars such as ours into the halo with velocities of 300-400~km~s$^{-1}$.

However, we do see several problems with this scenario. If the ejection was $\sim$25~Myr ago, when the star was near the disk, then the age of the putative cluster would have to be $\sim$155~Myr, given the 180~Myr age of the star.  Such clusters are not dense or compact; for comparison, 100~Myr is roughly the age of the Pleiades \citep{1993A&AS...98..477M}, which is neither.   More serious, however, is the issue of the lack of higher mass binaries to do the accelerating. If the age of the cluster is 155~Myr, we can expect the highest mass stars to be $\le$5$M_\odot$.  For a given orbital separation, the velocity of an ejected star will be only half as large if the binary has a mass of 5$M_\odot$+5$M_\odot$ rather than, say, 20$M_\odot$+20$M_\odot$.   The dynamical ejection mechanism is thus an unlikely explanation.

\subsubsection{Binary Supernova Scenario}

The classical explanation for the origin of OB runaways was first suggested by \citet{Zwicky} and subsequently by \citet{Blaauw}.   Two massive stars are in orbit about each other.  The more massive one evolves more quickly, and when it explodes as a core-collapse supernova, the system becomes unbound if enough mass is lost from the system.
The companion star then travels off at something like its tangential orbital velocity. 

Given the age of the star when it was in the disk ($\sim 150$~Myr), any companion would have to be $\leq$5$M_\odot$, well below the cutoff for a core-collapse supernova.   However, there is an intriguing alternative possibility that might apply to our star: it could be the surviving donor star of a Type Ia event produced via the single-degenerate progenitor channel \citep{1973ApJ...186.1007W}.

In this scenario Case A mass transfer in a binary would play a vital role. Imagine we began with a 3.1M$_{\odot}$+3.0$M_\odot$ binary. The 3.1$M_\odot$ primary expands first, overflows its Roche lobe, and dumps a significant amount of material onto the 3$M_\odot$ companion, increasing the companion mass to 5$M_\odot$ and leaving behind a white dwarf. The 5$M_\odot$ companion eventually expands as part of {\it its} evolution and undergoes a second bout of Roche lobe overflow mass transfer, transferring material onto the white dwarf. The white dwarf then explodes as a Type Ia supernova and the companion (our runaway star) is ejected.
 
 \citet{2001ApJ...552..664N} extensively explored Case A mass transfer parameter space; this scenario would be what they described as ``AL," which applies to binaries that start with relatively wide periods. In these cases the primary loses a lot of mass onto the secondary, and the secondary then fills its Roche Lobe and kicks off a ``reverse mass transfer." Their modeling stops right before the reverse mass transfer, but they note that in many of these scenarios the primary has become a white dwarf by the time this happens, so this is an excellent recipe for making a cataclysmic variable or a Type Ia supernova.  The  5.6~$M_\odot$ + 5.0$M_\odot$ simulation shown in their Figure 6 is similar to what we described.

There are caveats of course.  The separation between the stars needs to be in a ``sweet spot," large enough to have conservative mass loss rather than the formation of a common envelope.  Yet they also need to be close enough so that the (initial) secondary can overflow its Roche lobe back onto the primary (now a white dwarf) to form a Type Ia explosion. But, such an event would result with the surviving star to move off with a velocity compatible to the tangential orbital speed, with an extra ``kick" due to the explosion \citep{2000ApJS..128..615M,2001ApJ...550L..53C}.  

We realize that this binary scenario affects the presumed age of the star. However, the lifetime of a (single)  3$M_\odot$ star is still less than half a gigayear, and we would expect that the binary scenario would proceed along a time scale significantly shorter than this, with the
amount dependent upon the details of the orbital separation and mass ratio (see, e.g., \citealt{BPASS2}).

It is hard to evaluate the likelihood of such a scenario. No confirmed survivor of a Type~Ia supernova has ever been detected. While Tycho G has been proposed as a potential companion for the progenitor of SN 1572 \citep{2004Natur.431.1069R}, this result is still a matter of debate (e.g. \citealt{2013ApJ...765..150S, 2014MNRAS.439..354B}) and there has been no other discovery of a runaway Type Ia companion \citep{2014ARA&A..52..107M}. Predictions of these survivors' physical parameters have also focused on the star's appearance in the few thousand years following the supernova itself (e.g. \citealt{2000ApJS..128..615M, 2013ApJ...773...49P, 2013ApJ...765..150S, 2013A&A...554A.109L}), since prior searches have focused on identifying candidate companions for the small handful of recent (less than 1000 years old) Galactic Type~Ia supernova remnants. The expected long-term evolution and properties of a runaway companion from a Type Ia supernova are unknown. We therefore consider this an intriguing, if unsubstantiated, possibility.

We do note that the quest for clarifying the nature of the progenitor systems of Type~Ia supernovae is a subject of great interest (see \citealt{2014ARA&A..52..107M} for a recent review).  Do Type~Ia events come from double-degenerate or single-degenerate binaries, or do both channels contribute to Type~Ia events?  Thus, identification of a star as the survivor of a Type~Ia event would confirm a single-degenerate origin.  At the same time, recent {\it Gaia} data has identified three hypervelocity white dwarfs, lending credence to a double-degenerate origin \citep{2018arXiv180411163S}.

\subsection{Non-Disk Origins}
We note that this list of possible origins above is not exhaustive.  For instance, what if the progenitor cloud 
for this star was stripped off of a dwarf galaxy passing through the galactic halo $\sim$ 150 to 200~Myr ago? A condensing cloud could have been left on an orbit that gives us the star we see today, an explanation that avoids the problem of how to accelerate the star to its current velocity.  Of course, such a scenario is purely speculative, but the importance of mergers in galaxy evolution and the population of the halo is well established (see, e.g \citealt{2012ARA&A..50..251I}).   

Another possibility that we considered, but believe we can rule out, is that the star originated from the globular cluster NGC~362.  Our referee noted the intriguing coincidence that NGC~362 is located only half a degree away, and is at a similar distance \citep{NGC362}.  Furthermore, the {\it Gaia} broad-band color is quite similar to that of RGB stars in this cluster.   

However, we encountered several problems with this connection.  First, the cluster has an unusually high metallicity (for a globular cluster),  ${\rm Fe/H}=-1.26$ according to the 2011 update of the \citet{Harris96} catalog (see also \citealt{Carretta}, implying an unusually ``youthful"  age (for a globular cluster) of 10-11~Gyr.   However, both the metallicity and the age are well outside what is possible given the star's spectrum.  

To us though, the most convincing argument against a connection is kinematic.  Our star is located 5\farcm9 west
and 31\farcm4 south of the cluster, which has a radius of about 5\arcmin. (The cluster's core radius is 0\farcm18, \citealt{Harris96}).  Thus the star would have to somehow have been ejected from the cluster.  However, the star's proper motion is in the wrong direction to be outward bound: the cluster's proper motion is roughly 6.7~mas yr$^{-1}$ in right ascension and -2.6~mas yr$^{-1}$ in declination, while our star's proper motion is 8.65 and -0.91~mas yr$^{-1}$, respectively.   {\it Thus the star's relative motion is towards the cluster, not away from it, at a rate of 2.0~mas yr$^{-1}$ in right ascension and 1.7~mas yr$^{-1}$ in declination.}  Could the star be moving towards the cluster on a bound orbit?  Kinematically this does not work: we calculate a transverse velocity of 96.7 km s$^{-1}$ at the cluster's {\it Gaia} distance of 7.9~kpc.  The projected physical separation between the two is about 74~pc.  Even if the cluster's mass is as high as $10^6 M_\odot$, the escape velocity at a distance of 74~pc would be only 10.8~km~s$^{-1}$, and there is no possibility that the star is on a bound orbit.   So, we do not believe either the physical properties nor the kinematic measurements permit an origin in NGC~362.

\section{Summary}
\label{Sec-summary}
The star previously thought to be a runaway supergiant in the SMC has been shown by {\it Gaia} to be a member of the inner halo of our own Milky Way.  The {\it Gaia} proper motion, combined with our previous radial velocity measurement, shows that the star has an even higher velocity than previously thought, $\sim$300 km~s$^{-1}$, relative to the Galactic center, with most of the motion in the radial direction.  It is currently located 6.4~kpc below the Galactic plane.  

Our analysis of the physical properties of this G5-8~II has shown that the star has an effective temperature of 4800$\pm$100~K, and a luminosity $\log L/L_\odot$ of $2.70\pm0.20$.  Its location in the HRD is consistent with it being a $3-4M_\odot$ star on the red giant branch or early AGB phase. Its age is about 180~Myr.

The kinematics are consistent with the star having been ejected from the disk of the Milky Way about 25~Myr ago.  We discuss various ejection scenarios.  The most likely scenario is that the star was accelerated to its current location via the MBH in the center of the Milky Way.  The evidence for this is two-fold: first, the highly radial nature of the star's orbit argues for having been scattered to its current location from the center of the Galaxy, and secondly, analysis of the star's orbit shows that for some range of parallaxes between 8.9~kpc and 9.5~kpc the star would have passed exactly through the center.  Were the MBH explanation to be ruled out by improvements in the knowledge of its parallax, the next most viable explanation is that it was ejected as a result of a Type~Ia supernova event after undergoing Case A mass transfer in a binary system.  Dynamical ejection from a cluster seems unlikely given the expected age of the cluster (150~Myr) at the time of ejection.  We can rule out an origin in the adjacent NGC 362 globular cluster.  

Thus, the star J01020100-7122208 is in some ways even more extraordinary than originally proposed by \citet{YSGRunaway}.  Although HVS B stars in the halo are believed to have been ejected by the MBH, the argument has been mainly statistical \citet{2015ARA&A..53...15B}.  To our knowledge, this is the first case where an orbit analysis has demonstrated for a particular star that this ejection was quite likely, and has been made possible only by the extraordinary information being provided by the {\it Gaia} project. 

Whatever the explanation, it seems unlikely that J01020100-7122208 is a unique object. The original discovery of the star's extraordinary radial velocity by \citet{YSGRunaway} came about because the star happens to be seen in projection against the SMC.  It seems highly probable that {\it Gaia} data will enable the identification of many more such objects, of which J01020100-7122208 may serve as a prototype. Even a cursory examination of the {\it Gaia} DR2 shows other bright high velocity stars in the region of the SMC; however, spectroscopic followup is needed to establish their physical properties, such as temperature, metallicity, luminosity, mass, and age.  \citet{GaiaHVS} have already scrutinized the DR2 proper motions of a list of late-type halo stars that had previously suggested as HVS candidates based on ground-based large proper motions, which turned out to be over-estimated.  Although not HVSs, some of these stars may prove to be runaway intermediate-mass stars similar to J01020100-7122208, once their physical properties are determined.  Even more promising is the potential number of such stars that may be identified amongst the billions of stars that now have reliable distances, proper motions and radial velocities thanks to {\it Gaia.}

\acknowledgments

We are grateful for useful conversations with Henny Lamers and Paula Szkody, and for critical discussions with
Joe Llama, Deidre Hunter, and Cyril Georgy.  Comments by an anonymous referee on the first version of this paper were critical for setting us on the right path of our analysis.    This work has made use of data from the ESA
mission {\it Gaia} (\url{https://www.cosmos.esa.int/gaia}), processed by
the {\it Gaia} Data Processing and Analysis Consortium (DPAC,
\url{https://www.cosmos.esa.int/web/gaia/dpac/consortium}). Funding
for the DPAC has been provided by national institutions, in particular
the institutions participating in the {\it Gaia} Multilateral Agreement. We are grateful to the Carnegie and Arizona Time Allocation committees for their support, and to the excellent support we always receive observing on Las Campanas.

\facilities{Gaia, Magellan: Baade (MagE), Du Pont, Blanco (Hydra)}

\bibliographystyle{apj}
\bibliography{masterbib}

\begin{figure}
\epsscale{1.0}
\plotone{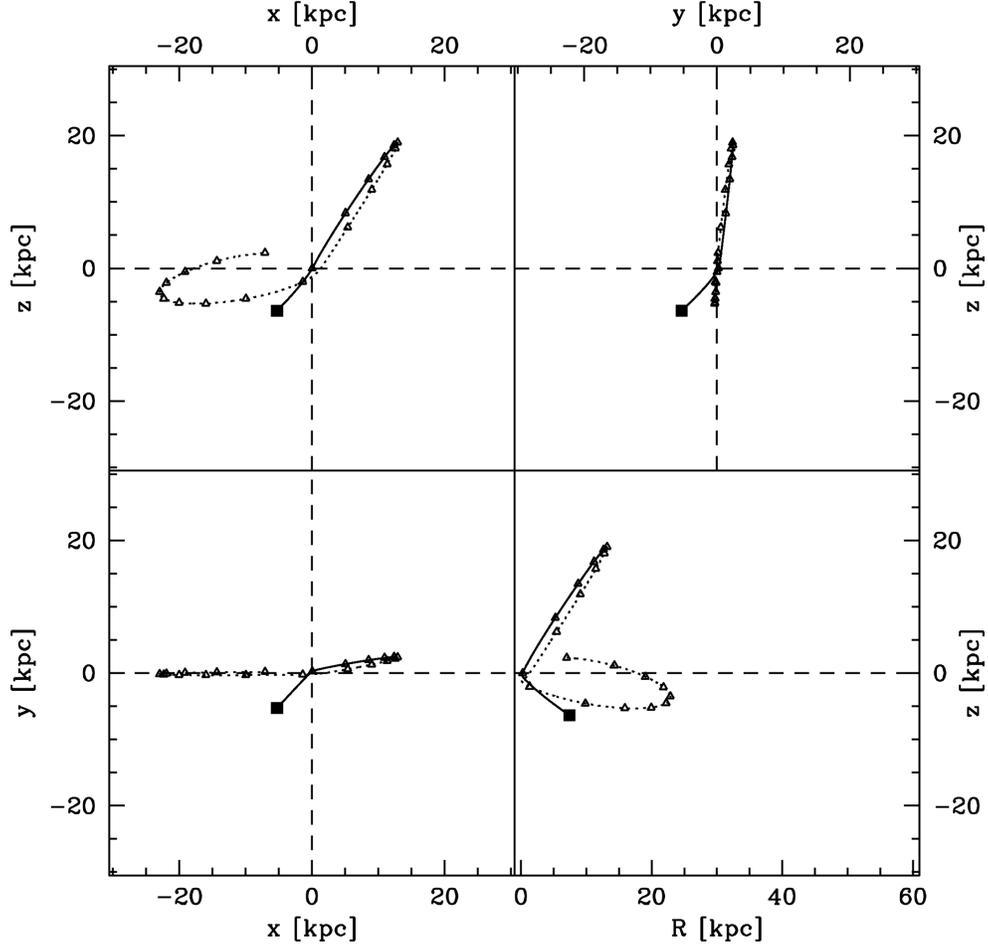}
\caption{\label{fig:SEL} The orbit of J01020100-7122208.  The orbit shown is for the third model discussed in the text, which we consider to be the most realistic, with \citet{1995A&A...300..117D} mass distribution. The black square shows the location of the star today (i.e., similar to Figure~\ref{fig:loc}), with the open triangles showing its positions in the past at intervals of 25~Myr. The dotted part of the orbit goes back in time beyond the time the star was born.}
\end{figure}

\begin{figure}
\epsscale{1.0}
\plottwo{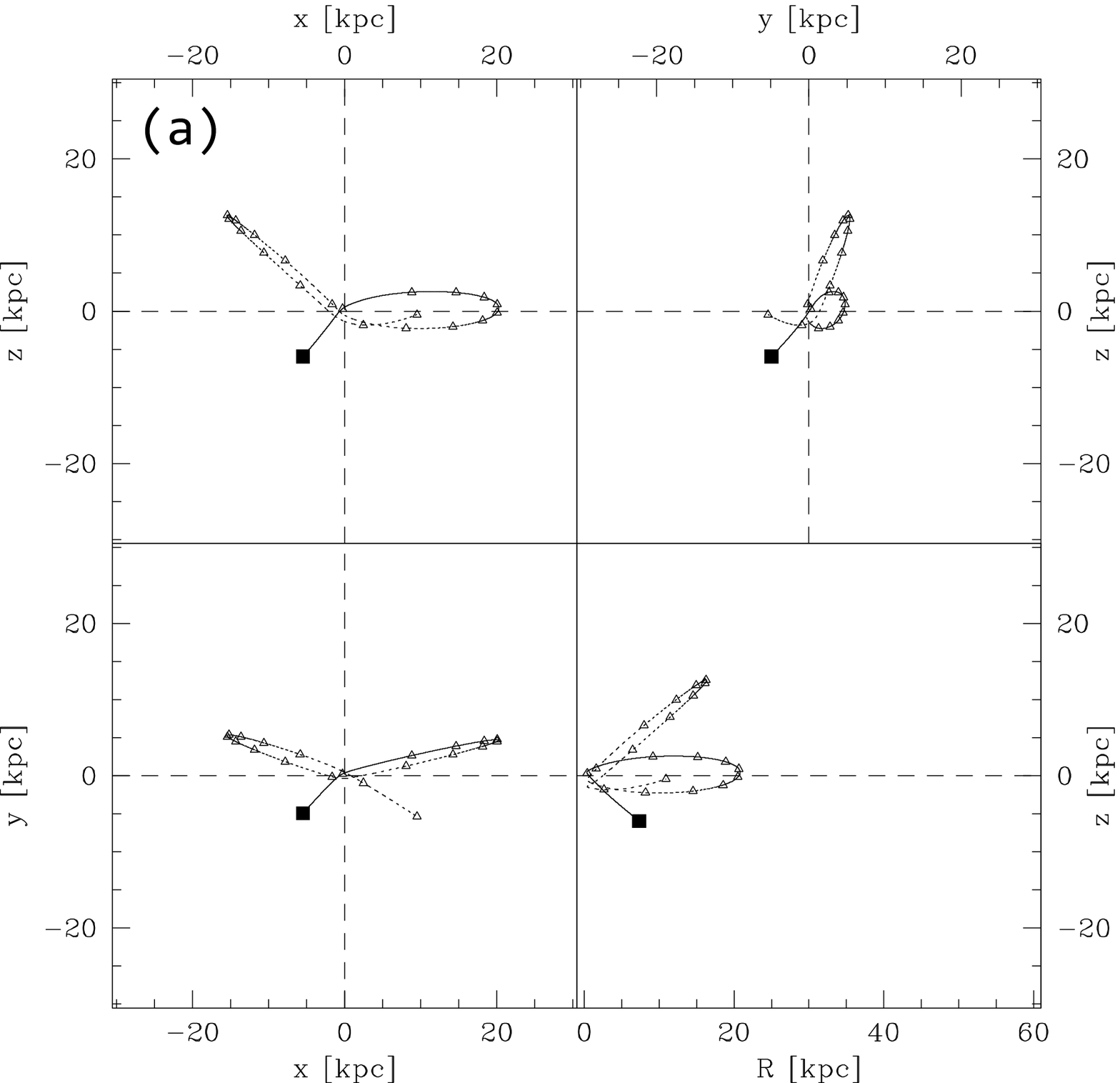}{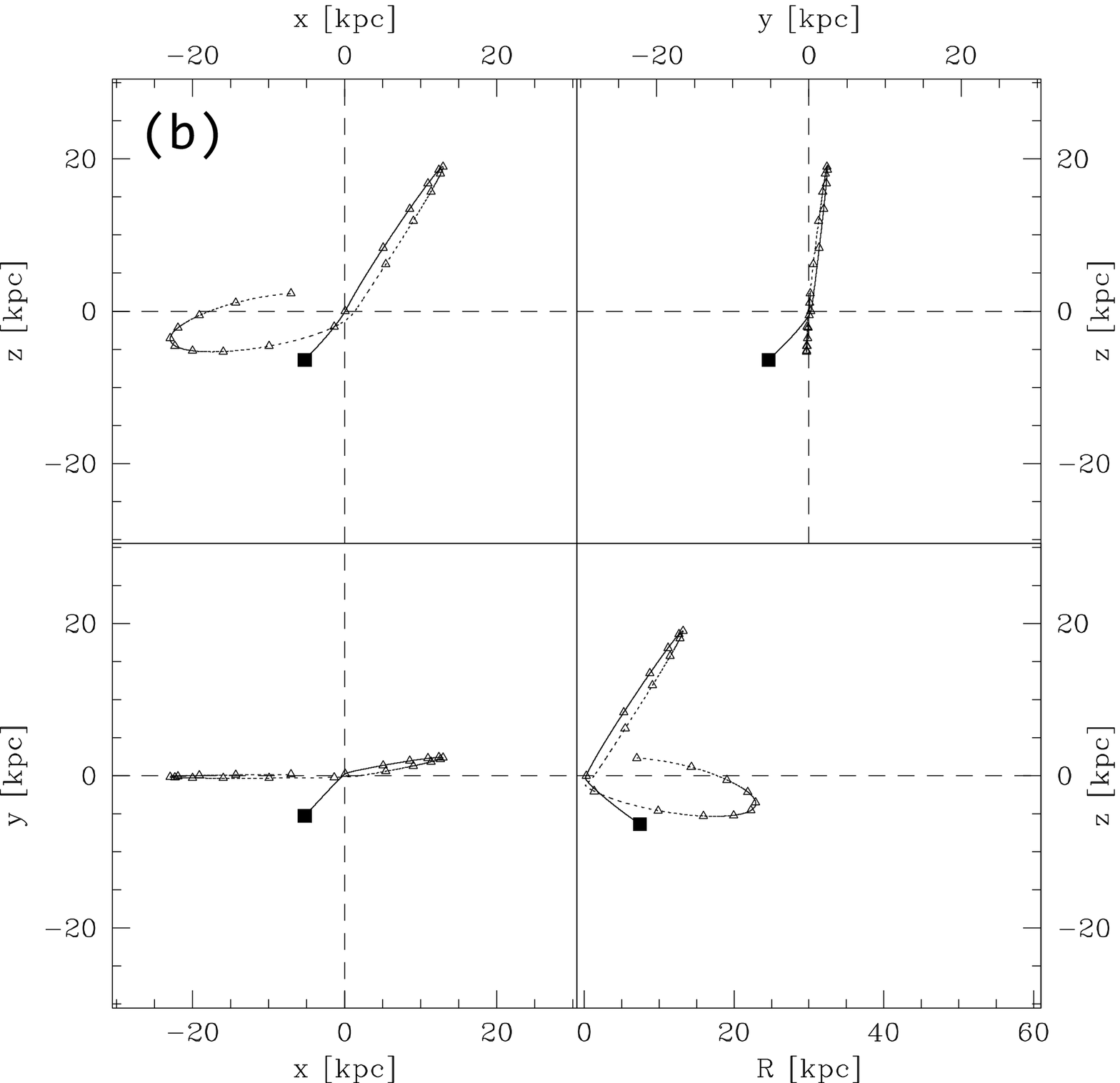}
\epsscale{0.5}
\plotone{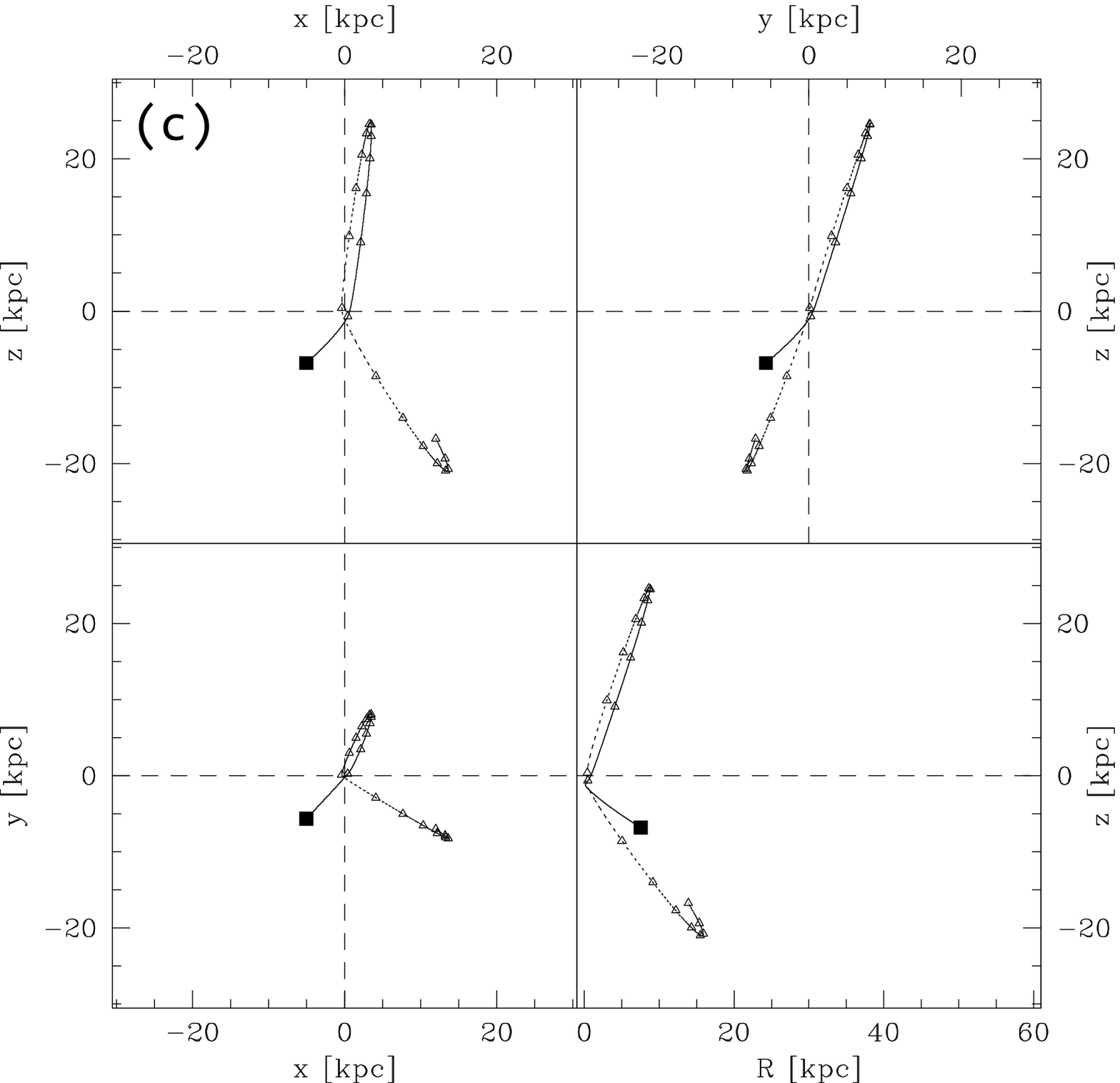}
\caption{\label{fig:rot} The orbit of J01020100-7122208 for three different parallaxes. Here we compare the orbit solutions for three different parallaxes, corresponding to distances of (a) 8.3~kpc, (b) our baseline of 8.9~kpc (i.e., identical to Figure~\ref{fig:SEL}), and (c) 9.5~kpc. The spacing and symbols are the same as in Figure~\ref{fig:SEL}. Note that sense of the rotation changes between the latter two values, demonstrating that for some parallax between 8.9~kpc and 9.5~kpc the star must pass through the exact center of the Galaxy.  Thus there is a range of parallaxes (possible a small range, but some range) where the star would have a sufficiently close passage with the MBH to scatter into the star into the halo.}
\end{figure}
\clearpage


\begin{deluxetable}{l C C C C C C C C C C C C}
 \tabletypesize{\scriptsize}
\rotate
\tablecaption{\label{tab:red} Intrinsic Colors and  Extinction from ATLAS9 [Fe/H]=-0.5 Models for $R_V=3.1$}
 \tablewidth{0pt}
 \tablehead{ 
 && \multicolumn{3}{c}{$T_{\rm eff}=4700$ } 
 &&\multicolumn{3}{c}{$T_{\rm eff}=4800$}
 &&\multicolumn{3}{c}{$T_{\rm eff}=4900$} \\ \cline{3-5} \cline{7-9} \cline{11-13}
 \colhead{Color} 
 & \colhead{Obs.\ Color}
 & \colhead{Int.\ Color}
 & \colhead{Color Excess}
 &\colhead{$A_V$}
 &
 & \colhead{Int.\ Color}
 & \colhead{Color Excess}
 &\colhead{$A_V$}
 &
 & \colhead{Int.\ Color}
 & \colhead{Color Excess}
 &\colhead{$A_V$}
 }
 \startdata
 U-B & 0.78\pm0.06 & 0.71 & 0.07 &  0.31\pm0.08 && 0.63 & 0.15 & 0.65\pm0.08 && 0.56 & 0.22 & 0.96\pm0.08 \\
 B-V &  1.15\pm0.05 & 1.06 & 0.09 & 0.28\pm0.05 && 1.01 & 0.14 & 0.43\pm0.05 && 0.96 & 0.19 & 0.57\pm0.05 \\
 V-K  & 2.93\pm0.06 & 2.44 & 0.49 & 0.52\pm0.02 && 2.33 & 0.60 & 0.63 \pm0.02 && 2.23 & 0.70 & 0.74\pm0.02 \\
 J-K  & 0.78\pm0.05 & 0.66 & 0.12 & 0.68\pm0.09 && 0.63 & 0.15 & 0.88\pm0.09 && 0.60 & 0.18 & 1.06\pm0.09 \\
 Wt.\ Avg.  & \nodata & \nodata & \nodata & $0.49\pm0.02$ && \nodata & \nodata & 0.62\pm0.02 & & \nodata & \nodata &0.74\pm0.02  \\
 \enddata
\end{deluxetable}


\begin{deluxetable}{l l}
\tablecaption{\label{tab:physical} Physical Properties}
\tablewidth{0pt}
\tablehead{
\colhead{Parameter}
&\colhead{Value}
}
\startdata
M$_{\rm V}$ & $-1.7\pm0.5$\tablenotemark{a}\\
T$_{\rm eff}$ & 4800$\pm$100 K\\
$\log \frac{{\rm L}}{{\rm L}_{\odot}}$ & $2.70\pm0.20$\tablenotemark{b}\\
Radius & $32\pm8$ R$_{\odot}$ \\
Mass & 3.5$\pm$0.5 M$_{\odot}$ \\
$\log$ g [cgs] & $2.0\pm0.2$\\
Age & 180 Myr  (130-300~Myr)\\
\enddata
\tablenotetext{a}{$-1.7\pm0.9$ if the full potential zero-point error of the {\it Gaia} parallax is included.}
\tablenotetext{b}{$2.70\pm0.56$ if the full potential zero-point error of the {\it Gaia} parallax is included.}
\end{deluxetable}

\begin{deluxetable}{l c}
\tablecaption{\label{tab:kinematics} Location and Kinematics}
\tablewidth{0pt}
\tablehead{
\colhead{Parameters}
&\colhead{Values}
}
\startdata
Celestrial coords.\ $\alpha_{\rm 2000}$, $\delta_{\rm 2000}$ & 01:02:01.00, -71:22:20.8 \\
Galactic coords.\ $l$,$b$ [degrees] & 301.722, -45.731 \\
Proper motion $\mu_\alpha \cos \delta$, $\mu_\delta$ [mas yr$^{-1}$] & $+8.647\pm0.036$, $-0.906\pm0.027$\\
Proper motion $\mu_l$, $\mu_b$ [mas yr$^{-1}$] & $-8.579\pm0.036$, $+1.413\pm0.027$ \\
Heliocentric radial velocity [km s$^{-1}$]& $+301\pm2.4$ \\
Distance from sun [kpc] & $8.9^{+2.8}_{-1.8}$\\
Galactocentric Cartesian x,y,z [kpc] & $-5.23\pm0.73$, $-5.28\pm1.19$, $-6.37\pm1.43$  \\
Velocity w.r.t. LSR (U, V, W) [km s$^-1$]& $+165.0\pm1.6$, $-400.1\pm1.8$, $-166.7\pm1.9$ \\
Velocity w.r.t. Galactic Center (X, Y, Z) [km s$^{-1}$] & $-165.0\pm1.6$, $-180.1\pm1.8$, $-166.7\pm1.9$ \\
Radial distance from the Galactic Center [kpc] & $9.8\pm1.2$\\
Radial velocity w.r.t Galactic Center [km s$^{-1}$] & $293.8\pm1.8$\\
Transverse velocity w.r.t. Galactic Center ($\theta$,$\phi$) [km s$^{-1}$] & $-32.2\pm1.8$, $+9.5\pm1.7$\\
Total space motion w.r.t. Galactic Center [km s$^{-1}$] & $295.7\pm3.1$ \\
\enddata
\end{deluxetable}

\end{document}